\input{aipcheck}

\documentclass[
    ,final            
  ,numberedheadings 
  ]
  {aipproc}

\layoutstyle{6x9}
\def\d{\mathrm{d}}
\usepackage{amsmath}
\usepackage{bm}

\begin{document}
\flushright
MAN/HEP/2011/17
\flushleft
\title{Phenomenology of Distribution Amplitudes for the $\rho$ meson}


\author{J. R. Forshaw}{
  address={School of Physics \& Astronomy, University of Manchester, \\
Oxford Road, Manchester M13 9PL, U.K.\\}
}

\author{R. Sandapen}{
  address={D\'epartement de Physique et d'Astronomie, Universit\'e de Moncton,
\\
Moncton, N-B. E1A 3E9, Canada.}
}

\begin{abstract}
We report on a successful extraction of the twist-$2$ and
twist-$3$ Distribution
Amplitudes (DAs) of the $\rho$ meson using the HERA data on diffractive $\rho$
photoproduction \cite{Forshaw:2011yj}. We extract these DAs using
several Colour Glass Condensate (CGC) inspired and a Regge inspired
dipole models. All our extracted DAs are consistent with Sum Rules and lattice predictions.

\end{abstract}

\maketitle


\section{Diffractive $\rho$ photoproduction}

In the dipole model \cite{Nikolaev:1990ja,Mueller:1994jq}, the 
imaginary part of the amplitude  for diffractive $\rho$ production is given
by~\cite{Marquet:2007qa}:
\begin{equation}
\Im \mbox{m} \mathcal{A}_{\lambda} (s,t;Q^2) =
\sum_{h,\bar{h}} \int \d^2  \mathbf{r} \; \d z \;
\Psi^{\gamma^*,\lambda}_{h,\bar{h}} (r,z;Q^2)
\Psi^{\rho,\lambda}_{h,\bar{h}}(r,z)^{*}
e^{-iz\mathbf{r} \cdot \mathrm{\Delta}}
\mathcal{N}(x,\mathbf{r},\mathrm{\Delta}) \;,
\label{non-forward-amplitude}
\end{equation}
where $t=-|\mathrm{\Delta}|^2$. $\Psi^{\gamma^*,\lambda}_{h,\bar{h}}(r,z;Q^2)$
and $\Psi^{\rho,\lambda}_{h,\bar{h}}(r,z)$ denote the light-cone wavefunctions
of the photon and $\rho$-meson respectively while
$\mathcal{N}(x,\mathbf{r},\mathrm{\Delta})$ is the universal dipole-proton
scattering
amplitude. The forward dipole-proton amplitude is well-constrained by the very
precise $F_2$ HERA data. The photon's light-cone wavefunctions are also
well-known, at least for large $Q^2$. We thus have an opportunity to
extract information on the meson's wavefunction using the HERA data
\cite{Chekanov:2007zr,Collaboration:2009xp} on diffractive $\rho$ production. To
do so, we use three CGC inspired and a Regge inspired dipole models. We refer to
them
as CGC[$0.63$], CGC[$0.74$], FSSat and t-CGC. The number in brackets for the
first two models refers to the value of the anomalous dimension $\gamma_s$. 
FSSat is the
Regge inspired model  that takes into account
saturation \cite{Forshaw:2004vv}. All of these models fit the $F_2$
data \cite{Forshaw:2004vv,Watt:2007nr} and also give a good description of the
diffractive structure function,
i.e.
$F_2^D(3)$ data \cite{Marquet:2007nf,Forshaw:2006np}. Finally, the t-CGC
\cite{Marquet:2007qa} model is the non-forward extension of the CGC[$0.74$]
model \cite{Soyez:2007kg}. 
  
\section{Wavefunctions and Distribution Amplitudes}

Previous work~\cite{Marquet:2007qa,Watt:2007nr,Forshaw:2003ki,Flensburg:2008ag}
has argued that a reasonable assumption for the
scalar part~\cite{Forshaw:2003ki} of the light-cone wavefunction for the $\rho$
is of the form
\begin{eqnarray}
\phi^{{\mathrm{BG}}}_\lambda(r,z) &=&
\mathcal{N}_\lambda \;  4[z(1-z)]^{b_{\lambda}} \sqrt{2\pi R_{\lambda}^{2}} \;
\exp \left(\frac{m_f^{2}R_{\lambda}^{2}}{2}\right)
\exp \left(-\frac{m_f^{2}R_{\lambda}^{2}}{8[z(1-z)]^{b_{\lambda}}}\right) \\
\nonumber
& &\times \exp \left(-\frac{2[z(1-z)]^{b_\lambda}
r^{2}}{R_{\lambda}^{2}}\right) \;,
\label{boosted-gaussian} 
\end{eqnarray}
which is referred to as the ``Boosted Gaussian'' (BG)
wavefunction.\footnote{This is a simplified version of the wavefunction
proposed in \cite{Nemchik:1996cw}.} In its
original form, $b_{\lambda}=1$ and $R_{\lambda}^2=12.9~\mbox{GeV}^{-2}$ so that
the leptonic decay width and normalization constraints are
satisfied~\cite{Forshaw:2003ki}. However, when this BG
wavefunction is used in conjunction with either the FSSat or any of the CGC
models, none of them
is able to give a good quantitative agreement with the current HERA data. The
situation improves considerably when we allow $R_\lambda$ and $b_\lambda$ to
vary freely. This results in an enhancement of the end-point
contributions~\cite{Forshaw:2011yj,Forshaw:2010py}. We also investigate the
requirement for additional end-point enhancement
in the
transverse wavefunction by using a
scalar wavefunction of the form
\begin{equation}
\phi_T (r,z)= \phi^{{\mathrm{BG}}}_T (r,z) \times [1+ c_{T}
\xi^2 + d_{T} \xi^4]
\label{EG} 
\end{equation}
where $\xi=2z-1$.
We fit to the 
total cross-section data, the
ratio of longitudinal to transverse cross-section data and the
decay constant datum for the longitudinally polarised meson, i.e. to a total of
$76$ data points. For the t-CGC model, we also include
the differential cross-section data with $|t| \le 0.5~\mbox{GeV}^2$ ($46$ data
points) resulting in a total of $122$ data points. Our best fit parameters
using each dipole model can be found in reference \cite{Forshaw:2011yj}.

Further constraints on the meson wavefunctions come from QCD Sum Rules and the
lattice which predict the moments of the corresponding Distribution
Amplitudes. These DAs parametrize the vacuum-to-meson transition matrix
elements of quark-antiquark non-local gauge invariant operators at light-like
separations \cite{Ball:2007zt}. In reference \cite{Forshaw:2011yj} we show that
the twist-$2$ and twist-$3$ DAs are related to the
scalar parts of the longitudinal and transverse light-cone wavefunctions
respectively. Explicitly the twist-$2$ DA 
\begin{equation}
\phi_\parallel(z,\mu) =\frac{N_c}{\pi \sqrt{2} f_{\rho} M_{\rho}} \int \d
r \mu
J_1(\mu r) [M_{\rho}^2 z(1-z) + m_f^2 -\nabla_r^2] \frac{\phi_L(r,z)}{z(1-z)}
\label{DA-scalar-lcwf-r-space-L}
\end{equation}
and the twist-$3$ DA
\begin{equation}
g_\perp(z,\mu)=\frac{N_c}{2 \pi \sqrt{2} f_{\rho} M_{\rho}} \int \d r \mu
J_1(\mu r)
\left[ (m_f^2 - (z^2+(1-z)^2) \nabla_r^2 \right] \frac{\phi_T(r,z)}{z^2 (1-z)^2
}\;.
\label{DA-scalar-lcwf-r-space-T} 
\end{equation}

The form of the DA at a low scale $\mu=1$ GeV can be constrained
using QCD Sum Rules \cite{Ball:2007zt}. It can then be evolved perturbatively to
any scale $\mu > 1$ GeV \cite{Ball:2007zt}. Before comparing to the Sum Rules
DAs, we must note that 
our extracted DA hardly evolves with the scale $\mu$ when $\mu \ge 1$ GeV, i.e.
it neglects the perturbatively known $\mu$-dependence. Given the limited $Q^2$
range of the HERA data to which we fit ($\sqrt{Q^2} < 7~$ GeV), our extracted DA
should thus be viewed as a parametrization at some low scale $\mu \sim 1$ GeV
\cite{Forshaw:2011yj,Forshaw:2010py}. 

\begin{figure}
\includegraphics[height=6cm]{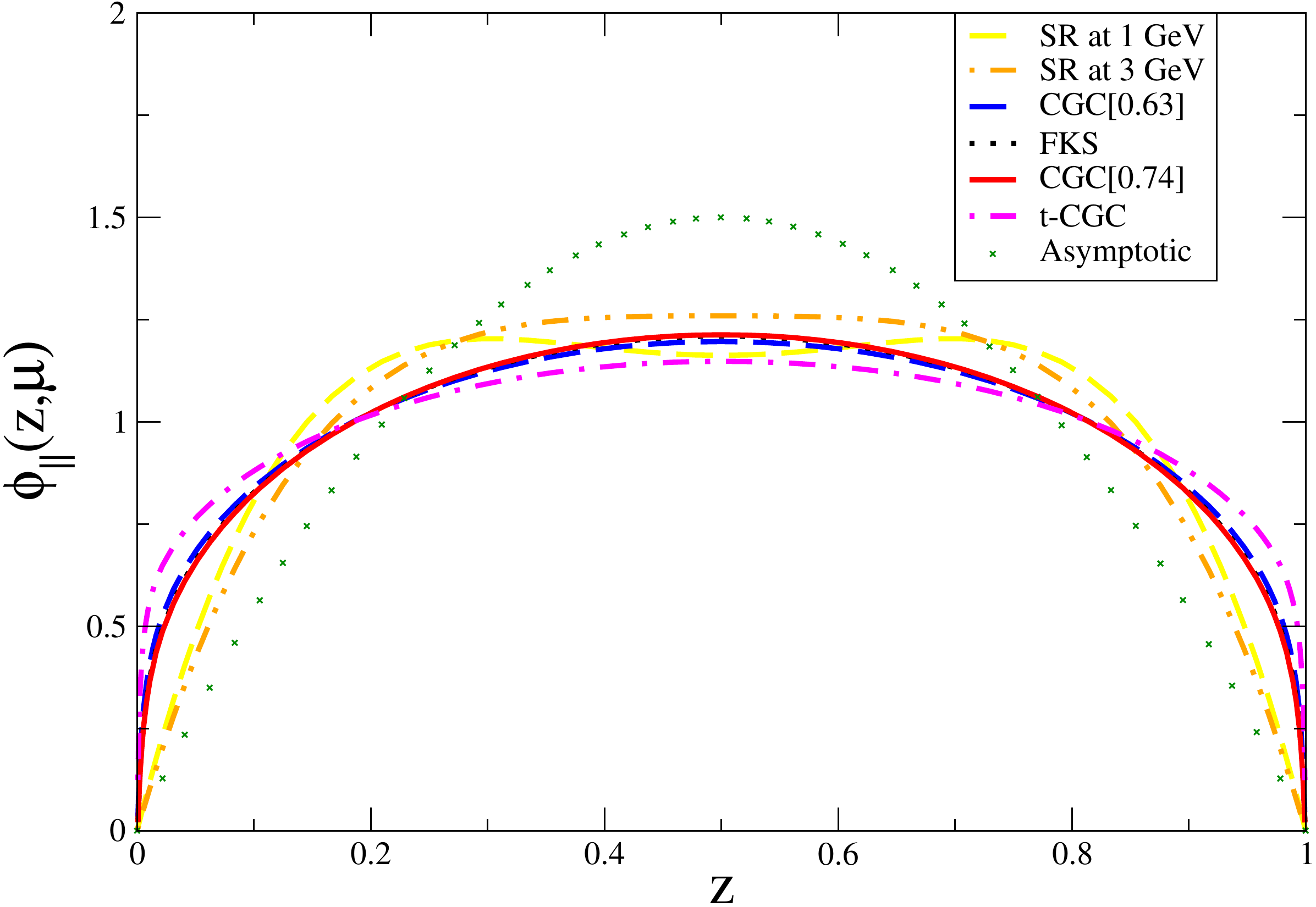} 
  \caption{The extracted leading twist-$2$ DAs compared to the
Sum Rule predictions. Dotted: FSSat; Solid: CGC[$0.74$];  Dashed: CGC[$0.63$];
Dot-dashed: t-CGC; Long-dashed: Sum Rules at $1$ GeV;
Dot-dot-dashed: Sum Rules at $3$ GeV. Crosses: Asymptotic.}
\label{fig:DAL}
\end{figure}

In figure \ref{fig:DAL}, we  compare our extracted leading twist-$2$ DAs to the
Sum Rules predictions. As can be seen, the extracted DAs
are all
broadly consistent
with the QCD Sum Rule distribution at $1$ or $3$ GeV. Note that the
non-monotonic behaviour of the Sum Rule
distribution is not physical and is due to the truncation in a
Gegenbauer expansion. 
Also shown is the narrower
asymptotic
distribution $\phi_{\parallel}(z,\infty)=6z(1-z)$. We also compute the second
moments of the leading twist DAs and found them to be consistent with Sum Rules
as well as lattice predictions \cite{Forshaw:2011yj}. 

\begin{figure}
\includegraphics[height=6cm]{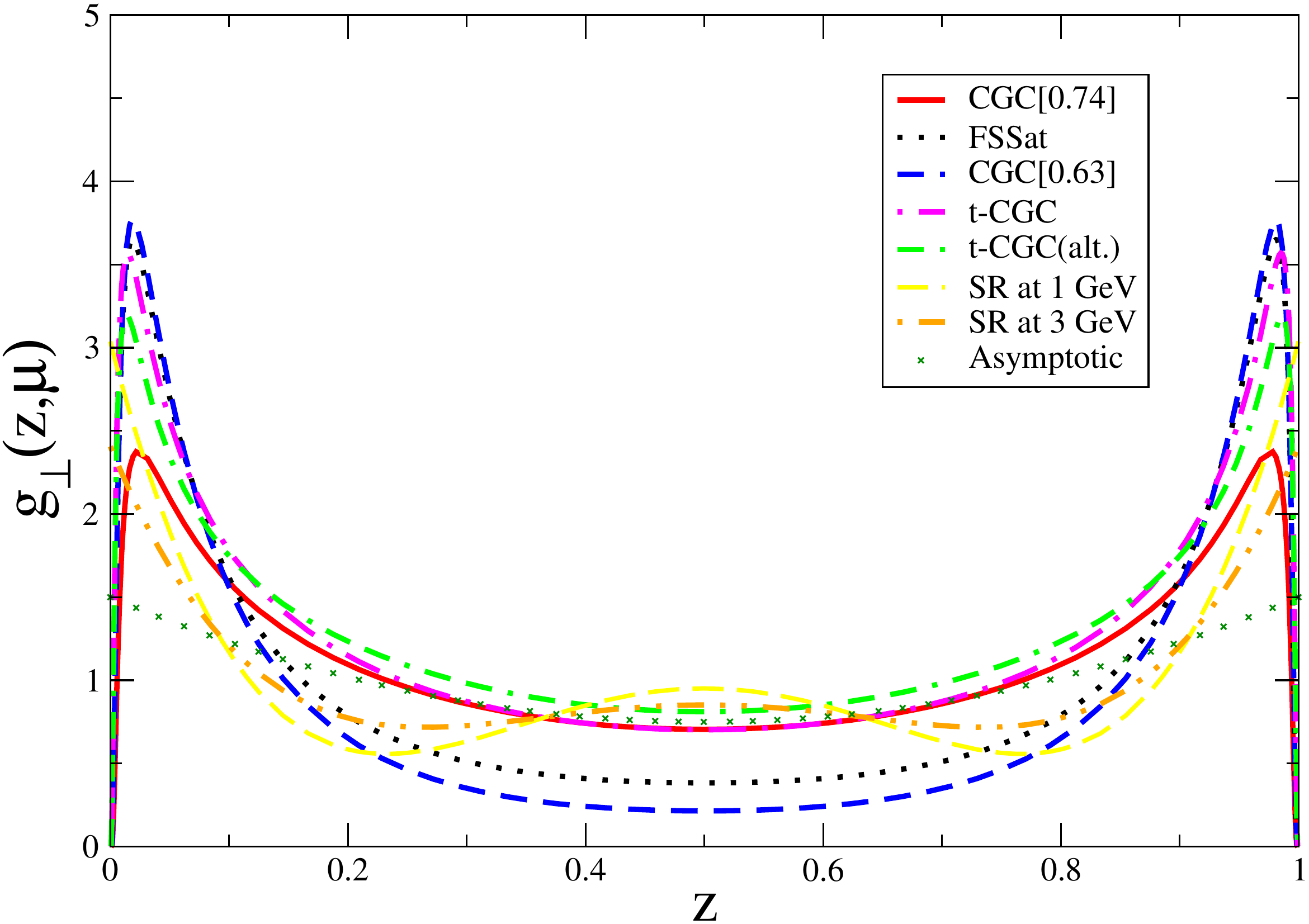}
\caption{The extracted twist-$3$ DAs compared to the QCD
Sum Rules predictions. Dotted: FSSat; Solid: CGC[$0.74$];  Dashed: CGC[$0.63$];
Dot-dashed: t-CGC; Dash-dash-dotted: t-CGC (alt.); Long-dashed: Sum Rules at $1$
GeV;
Dot-dot-dashed: Sum Rules at $3$ GeV. Crosses: Asymptotic.}
\label{fig:DAT}
\end{figure}

We compare in figure \ref{fig:DAT} our extracted twist-$3$ DAs with the Sum Rule
DA at two values of $\mu$: $1$ and $3$ GeV.  All the distributions show an
enhanced end-point contribution
compared to the asymptotic DA $g_{\perp}(z,\infty)=3/4(1+\xi^2)$ which is
also shown on the plot. However the degree of end-point enhancement is clearly
model-dependent. 


\section{Conclusions}
We have extracted the twist-$2$ and twist-$3$
Distribution Amplitudes of the $\rho$ meson using the HERA
data. We find that the twist-$2$ DA is not much model dependent whereas the data allow for a family of extracted twist-$3$ DAs with a varying 
degree of end-point enhancement. All our extracted DAs are consistent with Sum Rules and lattice predictions. 

In its present form, our DA lacks the peturbative evolution with the scale
$\mu$. If this is taken into account, a more precise comparison with the Sum
Rules and lattice predictions could be made.


\begin{theacknowledgments}
R.S. thanks  the organisers for a very enjoyable conference and acknowledges
financial support from the Facult\'e des Sciences 
and the Facult\'e d'\'Etudes Superieures et de la Recherche (FESR) of the Universit\'e de Moncton. This
research is also supported by the UK's STFC. 
\end{theacknowledgments}



\bibliographystyle{aipproc}   

\bibliography{sandapen}

\IfFileExists{\jobname.bbl}{}
 {\typeout{}
  \typeout{******************************************}
  \typeout{** Please run "bibtex \jobname" to optain}
  \typeout{** the bibliography and then re-run LaTeX}
  \typeout{** twice to fix the references!}
  \typeout{******************************************}
  \typeout{}
 }

\end{document}